\documentclass[sigconf,nonacm]{acmart}

\AtBeginDocument{%
  \providecommand\BibTeX{{%
    \normalfont B\kern-0.5em{\scshape i\kern-0.25em b}\kern-0.8em\TeX}}}

\copyrightyear{2022}
\acmYear{2022}
\setcopyright{acmlicensed}
\acmConference[ISSAC '22]{Proceedings of the 2022
International Symposium on Symbolic and Algebraic Computation}{July 4--7,
2022}{Villeneuve-d'Ascq, France}
\acmBooktitle{Proceedings of the 2022 International Symposium on Symbolic
and Algebraic Computation (ISSAC '22), July 4--7, 2022, Villeneuve-d'Ascq,
France}
\acmPrice{15.00}
\acmDOI{10.1145/3476446.3535498}
\acmISBN{978-1-4503-8688-3/22/07}

\AtEndPreamble{%
\theoremstyle{acmdefinition}
\newtheorem{remark}[theorem]{Remark}}

\usepackage{xspace}
\usepackage{bm}
\usepackage{tikz}

\newcommand{\R}{\mathbb{R}}
\renewcommand{\C}{\mathbb{C}}
\newcommand{\T}{(\C^*)^n}
\def\Pr{\mathbb{P}}
\def\Z{\mathbb{Z}}
\def\N{\mathbb{N}}

\def\P{\mathbb{P}}

\let\*\cdot

\def\grobner{Gr\"{o}bner\xspace}
\def\Groebner{\grobner}

\def\wrt{with respect to\xspace}
\def\gb{\grobner basis\xspace}
\def\gbs{\grobner bases\xspace}
\def\x{\bm{x}}
\def\NP{\mathtt{NP}}

\newcommand{\generators}[1]{\langle #1 \rangle}
\newcommand{\Supp}{\mathtt{Supp}}
\newcommand{\f}{\tilde{f}}
\newcommand{\MV}{\mathtt{MV}}
\newcommand{\Res}{\mathtt{Res}}

\begin{document}

\title{Solving sparse polynomial systems using Gröbner bases and resultants}

\author{Mat\'ias R. Bender}
\affiliation{%
  \institution{Technische Universität Berlin}
  \city{Berlin}
  \country{Germany}
}
\email{mbender@math.tu-berlin.de}
\orcid{0000-0001-9341-287X}

\begin{abstract}
  Solving systems of polynomial equations is a central problem in
  nonlinear and computational algebra. Since Buchberger's algorithm
  for computing Gröbner bases in the 60s, there has been a lot of
  progress in this domain. Moreover, these equations have been
  employed to model and solve problems from diverse disciplines such
  as biology, cryptography, and robotics. Currently, we have a good
  understanding of how to solve generic systems from a theoretical and
  algorithmic point of view. However, polynomial equations encountered
  in practice are usually structured, and so many properties and
  results about generic systems do not apply to them. For this reason,
  a common trend in the last decades has been to develop mathematical
  and algorithmic frameworks to exploit specific structures of systems
  of polynomials.

  Arguably, the most common structure is sparsity; that is, the
  polynomials of the systems only involve a few monomials. Since
  Bernstein, Khovanskii, and Kushnirenko's work on the expected number
  of solutions of sparse systems, toric geometry has been the default
  mathematical framework to employ sparsity. In particular, it is the
  crux of the matter behind the extension of classical tools to
  systems, such as resultant computations, homotopy continuation
  methods, and most recently, Gröbner bases. In this work, we will
  review these classical tools, their extensions, and recent progress
  in exploiting sparsity for solving polynomial systems.

  This manuscript complements its homonymous tutorial presented at the
  conference ISSAC 2022.
\end{abstract}



 \begin{teaserfigure}
   \vspace{-.5\baselineskip}
   \begin{center} \LARGE
     \emph{Dedicated to the memory of Agnes Szanto}
   \end{center}
   \vspace{.5\baselineskip}
  \end{teaserfigure}

\maketitle

\section{Introduction}

Systems of polynomial equations give us one of the simplest and
general ways of dealing with non-linear objects. They are expressive
enough to encode algebraic varieties and effective enough to allow us
to compute with them.
They generalize two ubiquitous kinds of equations appearing in
Mathematics, linear equations and univariate polynomials. 
Solving polynomial equations, that is, to find exact or approximate
solutions for the equations of the systems, is one of the central
problems in this setting and their applications span several domains
in science and engineering
\cite{cox_applications_2020,breidingNonlinearAlgebraApplications2021}.

There are several tools to solve polynomial systems, 
e.g., geometric resolutions
\cite{heintz2000deformation,giustiGrobnerFreeAlternative2001,herrero2013affine},
\gbs
\cite{faugereEfficientComputationZerodimensional1993,faugereNewEfficientAlgorithm1999}\cite[Chp.~2]{kreuzerComputationalCommutativeAlgebra2000},
homotopy continuation
\cite{morgan1987computing,huberPolyhedralMethodSolving1995,burgisserConditionGeometryNumerical2013},
normal form algorithms
\cite{lazardResolutionSystemesEquations1981,auzingerEliminationAlgorithmComputation1988,mourrainNewCriterionNormal1999a,telen2020thesis},
resultants \cite{cannyGeneralisedCharacteristicPolynomials1990,emirisMonomialBasesPolynomial1994}\cite[Chp.~3]{coxUsingAlgebraicGeometry2005},
subdivisions \cite{cucker2008numerical,mourrain2009subdivision,tonellicuetothesis},
subresultants \cite{gonzalez-vegaSubresultantTheoryMultivariate1991,szantoSolvingOverdeterminedSystems2008,diochnos2009asymptotic}
and triangular decompositions \cite{chen2012algorithms,wang2020decomposition}. 
The aforementioned is a non-exhaustive list of methods and references;
the interested reader can find the latest developments on many of
these methods in the proceedings of the previous editions of ISSAC.
Some nice introductions to the theory of solving systems of
polynomial equations can be found in
\cite{kreuzerComputationalCommutativeAlgebra2000,coxUsingAlgebraicGeometry2005,dickensteinSolvingPolynomialEquations2005}.
In this manuscript, and in its associated tutorial, we focus on \gbs,
resultants, and normal forms algorithms.

In practice, the polynomial systems that we encounter are structured,
e.g., sparse polynomials in biology \cite{gatermann2002family} and
statistics \cite{drton_lectures_2009}, symmetric in cryptography
\cite{faugere2010computing}, and determinantal in optimization
\cite{huber2000pieri}.
However, the general-purpose strategies to solve polynomial systems
does not exploit this structure.
For this reason, in the last years, an important trend in polynomial
system solving is to improve these techniques for specific structures.
In this text, we will focus on the sparsity of the inputs.
Besides sparsity, other kinds of structures include, e.g.,
symmetry \cite{faugere2013grobner,bazan2020ideal}, determinantal
\cite{faugere2010computing,labahn2021homotopy}, black box evaluations
\cite{burgisser2020rigid}, and degenerations
\cite{burr_numerical_2020}.

\subsection{Sparsity}

Sparsity is arguably the most common structure appearing in practice.
Before giving the definition of sparsity, let us introduce some notation.

\paragraph{Notation} Given $\alpha \in \Z^n$, we define
$\x^\alpha := \prod_{i=1}^n x_i^{\alpha_i}$. The monomials
$\{\x^\alpha : \alpha \in \Z^n\}$ generate the $\C$-algebra of
\emph{Laurent polynomials} $\C[x_1^{\pm},\dots,x_n^{\pm}]$.
We will write our polynomials in $\C[x_1^{\pm},\dots,x_n^{\pm}]$ as a
sum of terms, i.e., $\sum_{\alpha \in \N^n} c_\alpha x^\alpha$, where
$c_\alpha \in \C$ is the coefficient of the monomial $\x^\alpha$ and
there are finite $\alpha \in \Z^n$ such that $c_\alpha \neq 0$.
The \emph{support} of a given polynomial $f$ is the finite set of
exponents with non-zero coefficients, i.e.,
$$\Supp\left(\sum_{\alpha \in \N^n} c_\alpha \x^\alpha\right) := \{\alpha \in \Z^n
: c_\alpha \neq 0\}.$$

\noindent We abuse notation and we also identify $\Supp(f)$ with a set
of the monomials with exponents in $\Supp(f)$.
In this text, we will only work with rational polytopes, i.e., its
vertices are integer points.
Given two polytopes $P_1$ and $P_2$, we define its Minkowski sum $P_1
+ P_2$ as their point-wise addition, i.e., $P_1+P_2 := \{\alpha +
\beta : \alpha \in P_1, \beta \in P_2\}$. For each $i
\in \N$, let $i \, P := P + (i-1) \,
P$, and $0 \, P = \{0\}$.
We denote by $\Delta_n$ the $n$-dimensional standard simplex, i.e., the
convex hull of $\{0,e_1,\dots,e_n\}
\! \subset \! \R^n$, where
$\{e_1,\dots,e_n\}$ is the canonical basis of $\R^n$.

Intuitively, a polynomial $f \in \C[x_1,\dots,x_n]$ is sparse
when $\Supp(f)$ is \emph{small}, e.g., $\Supp(f)$ has less elements
than $\Supp(g)$, where $g \in \C[x_1,\dots,x_n]$ is a generic
polynomial of the same degree as $f$.
However, in this text we will use a related, but different, notion of
sparsity given by the \emph{Newton polytope} of the polynomial.

\begin{definition}
  [Newton polytope]

  Given a polynomial $f$, its \emph{Newton polytope}
  $\NP(f) \subset \R^n$ is the convex hull of $\Supp(f)$ over $\R^n$.
\end{definition}

Observe that the Newton polytope of a generic polynomial in
$\C[x_1,\dots,x_n]$ of degree $d$ is $d \, \Delta_n$, i.e., the
$d$-dilation of the $n$-dimensional standard simplex.

\begin{example}
[Generalized eigenvalue problem]
\label{ex:gep}

For most generalized eigenvalue problems in $\C^{2 \times 2}$, we can
represent them as a system of polynomial equations in two variables
$\lambda,w$, where $(1,\lambda)$ denotes the generalized eigenvalue and
$v := (1,w)$, the eigenvector. For example, for the pencil $(A,B)$,
$A := \big(\begin{smallmatrix}
  1 & 2\\
  3 & 4
  \end{smallmatrix}\big)$,
  $B := \big(\begin{smallmatrix}
    3 & 4\\
    -2 & -4
  \end{smallmatrix}\big)$, 
  we obtain the polynomial system $f_1 = f_2 = 0$,
  \begin{align*}
    (A + \lambda \, B) \, v = 0
    \iff
    \left\{
    \begin{array}{c}
    f_1 := 1 + 3 \, \lambda + 2 \, w + 4 \, \lambda \, w = 0  \\
    f_2 := 3 - 2 \, \lambda + 4 \, w - 4 \, \lambda \, w
    =
    0
    \end{array}
    \right. .
  \end{align*}
  The polynomials have degree 2 and their support are
  $\Supp(f_1) = \Supp(f_2) = \{1,\lambda,w,\lambda \, w\}$.
  The polynomials are sparse; written in terms of monomials of
  degree at most 2, we have that
  $$f_1 = \bm{1} + \bm{3} \, \lambda + \bm{0} \, \lambda^2 + \bm{2}
  \, w + \bm{0} \, w^2 + \bm{4} \, \lambda \, w.$$

  \noindent
  The gray square in the following picture is their Newton polytope.
  \begin{center}\small
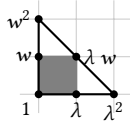
   \begin{tikzpicture}[scale=.5]
   \draw[step=1, black!20, thin] (-.5,-.5) grid (2.5, 2.5);
   \draw[thick,fill,color=gray] (0,0) -- (0,1) -- (1,1) -- (1,0) -- cycle;
   \draw[thick] (0,0) -- (0,2) -- (2,0) -- cycle;
   \fill (0,2) circle (3pt) node [left]  {$w^2$};
   \fill (0,1) circle (3pt) node [left]  {$w$};
   \fill (0,0) circle (3pt) node [below left]  {$1$};
   \fill (1,1) circle (3pt) node [right]  {$\lambda \, w$};
   \fill (1,0) circle (3pt) node [below]  {$\lambda$};
   \fill (2,0) circle (3pt) node [below]  {$\lambda^2$};
   \end{tikzpicture}
        \end{center}
\end{example}

\vspace{-.75\baselineskip} 

When we talk about sparsity, we look at Newton polytopes and not at
the supports because of the next theorem, attributed to Bernstein
\cite{bernshteinNumberRootsSystem1975}, Khovanskii
\cite{khovanskiiNewtonPolyhedraGenus1978}, and Kushnirenko
\cite{kushnirenkoNewtonPolytopesBezout1976}.

\begin{theorem}
  [BBK bound]

  Fix a polynomial system $(f_1,\dots,f_n)$ with a finite number of
  solutions over $\T$, where $\C^* := \C \setminus \{0\}$. Then, this
  number of solutions is upper bounded by the \emph{mixed volume} of
  $\NP(f_1),\dots,\NP(f_n)$, i.e., $\MV(\NP(f_1),\dots,\NP(f_n))$, where
  $$
  \MV(P_1,\dots,P_n) := 
  (-1)^n + \sum_{k = 1}^n (-1)^{n-k} \!\!\!\!\!\! \sum_{\substack{I \subset \{1,\dots,n\} \\ \#I = k}} \!\!\!\!\!\! \# \left( (P_{I_1} + \dots + P_{I_k}) \cap \Z^n \right) 
  $$

  Moreover, consider finite subsets $A_1,\dots,A_n \subset \Z^n$ and a
  system of generic polynomials $(f_1,\dots,f_n)$ supported on
  $A_1,\dots,A_n$, i.e., each
  $f_i = \sum_{\alpha \in A_i} c_{i,\alpha} \x^\alpha$, where
  $(c_{i,\alpha})_{i \in \{1,\dots,n\}, \alpha \in A_i}$ is a generic
  vector. Then, the aforementioned bound for this generic system is an
  equality.
\end{theorem}

For a proof and extensions of this theorem, see \cite{mondalHowManyZeroes2021}.
The generic conditions of the previous theorem can be relaxed
\cite{cannyOptimalConditionDetermining1991,chenUnmixingMixedVolume2019,BihanSoprunov_2019_unmixed}.

\begin{example}
  [{Cont. Ex.~\ref{ex:gep}}]
  
  Without considering the sparsity of the input, B\'ezout's theorem
  tells us that the system should have (at most) $2^2$
  solutions. However, it has only two, each corresponding to a
  different eigenspace. This number agrees with the mixed volume of
  the Newton polytopes defining the equations.
\end{example}

When working with sparse polynomials, we usually consider two class of
systems.  We say that a sparse polynomial system $(f_1,\dots,f_r)$ is
\emph{unmixed} when there is an integer polytope $P \subset \R^n$ and
integers $d_1,\dots,d_r$ such that $\NP(f_i) = d_i \, P$, i.e., each
Newton polytope is a dilation of a common integer polytope.
Otherwise, the system is \emph{mixed}.

In this manuscript, we will discuss \gbs and resultants in the context
of sparse polynomials. We will employ the Newton polytopes of the
input polynomials to speed up the computations and derive complexity
bounds for our algorithms depending on the sparsity pattern of the
inputs. These techniques rely on \emph{toric geometry}.
Toric geometry was also employed to solve sparse polynomial systems
using other techniques such as homotopy continuation
\cite{huberPolyhedralMethodSolving1995,malajovich2019complexity,ergurPolyhedralHomotopyAlgorithm2021}
or geometric resolutions \cite{herrero2013affine}.

\begin{remark}
  [Other notions of sparsity]

  As the BKK bound shows, from a complex point of view, the solutions
  of the sparse systems are ``determined'' by the Newton polytopes of
  the input polynomials and not by their supports. However, from a
  real point of view, the support matters. In real algebraic geometry,
  a polynomial whose support is small is called a
  \emph{fewnomial}. Khovanskii showed that the number of positive real
  solutions of a fewnomial system is upper bounded by a quantity
  singly exponential in the size of the supports of the fewnomials
  \cite{khovanskii1991fewnomials}.
  Determining the number of real solutions of a fewnomial system is an
  active area of research. For example, the bounds were improved for
  general \cite{bihan2007new} and particular systems
  \cite{bihan2017descartes}. Moreover, probabilistically, stronger
  bounds hold \cite{burgisser2019number,jindal2020many}.

  A different kind of sparsity that has been studied is the
  \emph{chordal structure}.  This structure only cares about the
  variables appearing in each polynomial, regardless of the specific
  monomials.
  There are dedicated algorithms to compute \gbs
  \cite{cifuentes2016exploiting} and triangular sets
  \cite{mou2021chordal} that exploit this notion.
\end{remark}

\subsection{Multiplication maps}



An intermediate tool used to solve polynomial systems using the \gbs and
resultants is multiplication maps. To define them,
let $\bm{f} := \{f_1,\dots,f_r\} \subset \C[x_1,\dots,x_n]$ be an
affine polynomial system with a finite number of solutions $\delta$
over $\C^n$, counting multiplicities. The quotient ring
$\C[x_1,\dots,x_n] / \generators{\bm{f}}$ is a finite dimensional
$\C$-vector space whose dimension is $\delta$
\cite[Chp. 4.2]{coxUsingAlgebraicGeometry2005}, so we will fix a basis
$b_1,\dots,b_\delta$ to identify
$\C[x_1,\dots,x_n] / \generators{\bm{f}}$ with $\C^\delta$.
Given $g \in \C[x_1,\dots,x_n] / \generators{\bm{f}}$, the
\emph{multiplication map} of $g$ is the morphism
\begin{align*}
M_g : \C[x_1,\dots,x_n] / \generators{\bm{f}} & \rightarrow
                                                       \C[x_1,\dots,x_n] / \generators{\bm{f}}   \\
  h & \mapsto M_g(h) := g \, h.
\end{align*}
Using the basis $\{b_1,\dots,b_\delta\}$, we think about $M_g$ as a
matrix, which we call the \emph{multiplication matrix}.
Multiplication matrices are ubiquitous in algorithms for solving
polynomial systems:
\begin{itemize}
\item On one hand, we can approximate the solutions by computing
  the eigenvalues of these maps. More precisely, the \emph{eigenvalue
    theorem} \cite{lazardResolutionSystemesEquations1981} states that
  the eigenvalues of $M_g$ correspond to the evaluations\footnote{The
    evaluation of $g \in \C[x_1,\dots,x_n] / \generators{\bm{f}}$ at a
    point $p \in \C^n$ corresponds to $\bar{g}(p)$, where
    $\bar{g} \in \C[x_1,\dots,x_n]$ is such that $\bar{g}$ belongs to
    the class of $g$ in $\C[x_1,\dots,x_n] / \generators{\bm{f}}$.} of $g$ at the $\delta$
  solutions of the system $(f_1,\dots,f_r)$. See
  \cite{coxStickelbergerEigenvalueTheorem2021} for more details on
  this theorem and its history. Moreover, for each solution $p$ of
  $\bm{f}$, the vector given by the evaluations
  $(b_i(p))_{i \leq \delta}$ belongs to the eigenspace of $M_g$ of
  eigenvalue $g(p)$
  \cite{auzingerEliminationAlgorithmComputation1988}.

\item On the other hand, multiplication maps are a standard tool to
  compute \gbs for zero-dimension ideals. The algorithm FGLM
  \cite{faugereEfficientComputationZerodimensional1993,faugereSparseFGLMAlgorithms2017}
  allows us to derive, from the multiplication matrices, \gbs of
  $\generators{\bm{f}}$ \wrt any monomial order, e.g.,
  lexicographical, or to find a rational univariate representation of
  the solutions \cite{rouillier1999solving}.
\end{itemize}

To compute multiplication matrices, we use a \emph{normal form}, which
is, roughly speaking, a homomorphism
$\C[x_1,\dots,x_n] \rightarrow \C^\delta$ whose kernel is
$\generators{\bm{f}}$.
We can obtain such normal forms using, for example, \gbs
\cite[Chp.~2]{kreuzerComputationalCommutativeAlgebra2000}, border bases
\cite{mourrainNewCriterionNormal1999a}\cite[Chp.~4]{dickensteinSolvingPolynomialEquations2005},
or Sylvester formulas for the resultant
\cite{auzingerEliminationAlgorithmComputation1988,telenSolvingPolynomialSystems2018,telen2020thesis}.
We can compute these matrices by performing $d^{O(n)}$ arithmetic
operations, where $d$ is the maximal among the degrees of the
polynomials in $\bm{f}$ \cite{lakshman1991complexity}.

The aforementioned strategies do not employ the sparsity of the inputs
to speed up the computations (at least
directly\footnote{Heuristically, implementations of the F4 algorithm
  to compute \gbs \cite{faugereNewEfficientAlgorithm1999} can benefit
  partially from the sparsity of the inputs; see, e.g.,
  \cite{monagan2017algorithm,berthomieu2021msolve}.}). Moreover, their
complexity bounds are independent of this structure and to achieve
them, we must destroy the sparsity of the inputs, i.e., to perform
random linear change of coordinates or to consider random combinations
of the input polynomials.
In the following sections, we discuss alternative approaches to avoid
these issues.

\section{The sparse resultant}

In this section, we will assume that the reader is familiar with the
classical (projective) resultant. We refer to
\cite[Chp.~3]{coxUsingAlgebraicGeometry2005} and
\cite{buse:inria-00077120} for an introduction.

The sparse resultant is a generalization of the classical resultant
that allow us to decide if a sparse affine polynomial system in $n$
variables given by $n+1$ polynomials has a common solution in
$(\C^*)^n$. It is one of the first available tools to solve sparse
polynomial systems and a subject of study since the 90's. 

Given finite sets $A_0,\dots,A_n \subset \Z^n$, we can
parameterize each different polynomial system supported in
$\bm{A} := (A_0,\dots,A_n)$ with points\footnote{We identify the systems using the
  multi-projective space because multiplying the polynomials by
  non-zero constants does not change their solution set.}  in
$\P^{\bm{A}} := \P^{\#A_0 - 1} \times \dots \times \P^{\#A_n - 1}$,
$$(\bm{c_0},\dots,\bm{c_n}) \in \P^{A_0} \times \dots \times \P^{A_n} \mapsto  \left(\sum_{\alpha \in A_0} c_{0,\alpha} x^\alpha,\dots,\sum_{\alpha \in A_n} c_{n,\alpha} x^\alpha\right).$$

\noindent
We use the \emph{incidence variety} $\Omega$ to characterize all the
sparse systems of polynomials supported on $A_0,\dots,A_n$ with
solutions in $\T$.
$$
\Omega := 
\{(p,(F_0,\dots,F_n)) \in (\C^*)^n \times \P^{\bm{A}} : (\forall i) \, F_i(p) = 0\},
$$

\noindent
If we consider the projection map
$\pi : (\C^*)^n \times \P^{\bm{A}} \rightarrow \P^{\bm{A}}$,
  $\pi(\Omega) \subset \P^{\bm{A}}$ determines the systems with
  solutions at $\T$. However, as $\pi(\Omega)$ is not an algebraic
  variety, we consider its algebraic closure $\overline{\pi(\Omega)}$ in
  $\P^{\bm{A}}$.
  Under combinatorial assumptions on $\bm{A}$
  \cite[Cor.~1.1]{sturmfelsNewtonPolytopeResultant1994},
  $\overline{\pi(\Omega)}$ has co-dimension $1$ on $\P^{\bm{A}}$ and
  it is an irreducible hypersurface defined by a multihomogeneous
  polynomial $\Res_{\bm{A}}$ called the \emph{sparse resultant} of
  $\bm{A}$\footnote{We follow Esterov's \cite{esterov2010newton} and
    D'Andrea \& Sombra's \cite{dandreaPoissonFormulaSparse2015}
    definition of the sparse resultant. That is, the polynomial
    $\Res_{\bm{A}}$ might not be irreducible, but it is the $d$-power
    of an irreducible polynomial, the \emph{eliminant}, where $d$ is
    the degree of $\pi$ restricted to $\Omega$; see
    \cite[Sec.~2]{dandreaPoissonFormulaSparse2015}.
    Classically 
    \cite{gelfandDiscriminantsResultantsMultidimensional2008,sturmfelsNewtonPolytopeResultant1994},
    the sparse resultant was defined as the eliminant.
  }.
    For each $i$, the sparse resultant $\Res_{\bm{A}}$ is homogeneous
    \wrt the coefficients of the polynomial $F_i$ supported in $A_i$
    and its degree is $\MV(P_0,\dots,P_{i-1},P_{i+1},\dots,P_n)$,
    where $P_i$ is the convex hull of $A_i$.
The sparse resultant is a generalization of the classical (projective)
resultant. More precisely, if we have that
$A_i = d_i \Delta_n \cap \Z^n$, for some $d_i \in \N$, for each $i$,
the sparse and classical resultant agree.
The interested reader can find more properties of the sparse resultant
in, e.g.,
\cite[Chp.~7]{coxUsingAlgebraicGeometry2005},\cite[Chp.~8]{gelfandDiscriminantsResultantsMultidimensional2008},
\cite{sturmfelsNewtonPolytopeResultant1994,dandreaPoissonFormulaSparse2015}.

\subsection{Vanishing of the sparse resultant}

As the classical resultant is a special case of the sparse resultant,
it is easy to verify that having solutions on the torus $\T$ is a
necessary but not sufficient condition for the vanishing of the sparse
resultant.
To understand when this vanishes, let us recall when the classical
(projective) resultant vanishes.
Given an affine polynomial $f \in \C[x_1,\dots,x_n]$, we define its
classical \emph{homogenization}, $f^h \in \C[x_0,x_1,\dots,x_n]$, as
the polynomial  $f^h = x_0^{\mathtt{deg}(f)} \, f
\left(\frac{x_1}{x_0},\dots,\frac{x_n}{x_0} \right)$.
Given an affine polynomial system
$f_0,\dots,f_n \in \C[x_1,\dots,x_n]$, the classical
resultant of $(f_0,\dots,f_n)$ will vanish if and only if the
\emph{homogenization} of the system, i.e.
$f_0^h,\dots,f_n^h \in \C[x_0,x_1,\dots,x_n]$, has a common solution
on $\P^n$.
The homogenization does not change the solutions of the original
system on $\C^n$, i.e., for $p \in \C^n$, if $f(p_1,\dots,p_n) = 0$,
then $f^h(1,p_1,\dots,p_n) = 0$.
However, it might introduce new solutions \emph{at infinity}, i.e.,
$\P^n \setminus \C^n := \{[p_0 : \dots : p_n] \in \P^n : p_0 = 0\}$.
In an analogous way, the sparse resultant $\Res_{\bm{A}}$ of a sparse
system will vanish if and only if a homogenization of the system
vanishes over a normal projective toric variety ${X}$.
This toric variety is a compact variety containing $\T$ which is
constructed from the Minkowski sum $P := \sum_i P_i$; see, e.g.,
\cite[Chp.~2.3]{coxToricVarieties2011} for more details on this
construction.
As in the classical setting, the homogenization of the system has the
same solutions as the original system over $\T$, but might have
extra zeros on $X \setminus \T$, which we will call \emph{zeros at
  infinity}.
More precisely, for each $P_i$, we can find a torus-invariant nef
Cartier divisor $D_i$ such that the global sections of its associated
line bundle, denoted by $O_X(D_i)$, can be identified with the sparse
polynomials whose Newton polytope is equal to (or contained in) $P_i$ \cite[Chp.~6]{coxToricVarieties2011},
that is,
\begin{align} \label{eq:globalSec}
H^0(X_P,O_X(D_i)) = \{g \in \C[x_1^{\pm},\dots,x_n^{\pm}] : \NP(g)
\subseteq P_i\}.
\end{align}
This identification corresponds to the homogenization. Somehow, if we
have a global section $s \in H^0(X_P,O_X(D))$ identified with
$g := \sum_{\alpha \in Q} c_\alpha x^\alpha$, then for every
$p \in \T$, $s(p) = 0$ if and only if $g(p) = 0$.
Using this intuition, we consider the incidence variety
$$
\Omega^h := \{(p,(F_0,\dots,F_n)) \in X_{\bm{P}} \times \P^{\bm{A}} :
(\forall i) \, \mathtt{hom}(F_i)(p) = 0\},
$$

\noindent
where $\mathtt{hom}(F_i)$ denotes the global section of $O_X(D)$
identified with $F_i$.
If $\pi$ now denotes the projection onto $\P^{\bm{A}}$,
$\pi(\Omega^h)$ is an algebraic variety (because $X$ is
projective) which we can prove agrees with
$\overline{\pi(\Omega)}$.
We refer the interested reader to
\cite[Chp.~7.3]{coxUsingAlgebraicGeometry2005} for a more detailed
explanation in the case of unmixed systems\footnote{Beware that the
  toric variety constructed in
  \cite[Chp.~7.3]{coxUsingAlgebraicGeometry2005} and $X$ might not
  agree ($X$ is its normalization; see
  \cite[Chp.~3.A]{coxToricVarieties2011}), but the argument extends
  to $X$.}.


\begin{remark}
  To define a homogenization in terms of coordinates, as we did
  over $\P^n$, we need to introduce a coordinate ring for
  $X$. However, such a ring depends on the way (embedding) that we use
  to think about $X$.
  For example, we can think about $X$ as an \emph{almost geometric
    quotient} and use its associated ring, the Cox ring
  \cite{coxHomogeneousCoordinateRing1992}, to homogenize the
  polynomials; see
  \cite[Chp.~5]{coxToricVarieties2011}. Alternatively, we can use the
  so-called Cayley trick
  \cite[Chp.~8]{gelfandDiscriminantsResultantsMultidimensional2008},
  where we choose nef line bundles on $X$ associated to each $P_i$ and
  consider the embedding of $X$ on the (weighted) multi-projective
  space defined by them.
  We will use the latter approach in Sec.~\ref{sec:gbs}.
\end{remark}

\vspace{-1\baselineskip}

\subsection{Computing the sparse resultant}

\subsubsection{The determinant of a matrix}

A classical way of computing the resultant is a factor of the
determinant of a matrix. Sylvester \cite{sylvester_1840} showed that
we can compute the resultant of two univariate polynomials
$f_0,f_1 \in \C[x_0,x_1]$ by considering the determinant of the
Sylvester matrix, which linearizes the (Sylvester) map
$(g_0,g_1) \mapsto g_0 \, f_0 + g_1 \, f_1$, where $g_0$ is a
polynomial of degree at most  $\deg(f_1) - 1$ and $g_1$ a polynomial of
degree at most $\deg(f_0) - 1$.
This construction was later generalized by Macaulay
\cite{macaulay1994algebraic}, who showed how to compute the classical
(projective) resultant of $n+1$ polynomials
$f_0,\dots,f_n \in \C[x_1,\dots,x_n]$ of degree $d_0,\dots,d_n$.
He also showed that there is a way of constructing monomial sets
$B_0,\dots,B_n$ such that each $x^\beta \in B_i$ has degree at most
$\sum_{j \neq i} d_j - n$ and $\sum_{i} \#B_i$ agrees with the number
of monomials of degree at most $\sum_{j} d_j - n$ in
$\C[x_1,\dots,x_n]$. With these considerations, we can construct the
\emph{Macaulay matrix} $M$, whose columns we index with monomials in
$\C[x_0,\dots,x_n]$ of degree at most $\sum_{j} d_j - n$ and the rows with
pairs $\{(i,x^\beta) : i \in \{0,\dots,n\}, x^\beta \in B_i\}$.
The element in $M$ associated with the column indexed by $x^\alpha$ and row
indexed by $(i,x^\beta)$ corresponds to the coefficient of the
monomial $x^\alpha$ in $x^\beta \, f_i$.
Macaulay proved that the resultant of $(f_0,\dots,f_n)$ is a factor of
the determinant of $M$, i.e., $\det(M) = \Res(f_0,\dots,f_n) \cdot
E$. Moreover, he showed that the polynomial $E$, usually called the
\emph{extra factor}, is a specific minor of $M$.
Observe that the matrix $M$ is the transpose of a matrix representing
the Sylvester map $(g_0,\dots,g_n) \mapsto \sum_i f_i \, g_i$, where
the support of each $g_i$ is contained in $B_i$. For this reason, we
say that his construction is a \emph{Sylvester formula} for the
resultant.
Besides Sylvester formulas, we observe that other maps can be used to compute resultants. We refer the reader to
\cite{emirisMatricesEliminationTheory1999} and references therein.

Canny and Emiris \cite{cannyEfficientAlgorithmSparse1993} generalized
Macaulay's construction to the sparse case. They presented a Sylvester
formula for the resultant constructed out of mixed subdivisions of the
Newton polytopes of the input polynomials.
The interested reader can find more details and references in
\cite[Chp.~7.6]{coxUsingAlgebraicGeometry2005}.
However, Canny and Emiris did not characterize the extra factor
appearing in their constructions.
Different authors studied this question, e.g.,
\cite{d2002macaulay,groh2020subdivisions,d2022canny}, and very
recently D'Andrea, Jeronimo, and Sombra \cite{d2022canny} presented a
complete characterization of the extra factor which allowed them to
recover Macaulay's original results.

\vspace{-.5\baselineskip}
\subsubsection{The determinant of a complex}
A more general and elaborated way of computing the resultant is as the
determinant of a complex of vector spaces. This approach is attributed
to Cayley and was extended to study general kinds of resultants, such as the
sparse one; the interested reader can find more details in
\cite[Chp.~3.4]{gelfandDiscriminantsResultantsMultidimensional2008}.
It turns out that the aforementioned Sylvester maps are just the last
maps appearing in certain Koszul complexes that we can use to compute
resultants; see, e.g.,
\cite[Chp.~13]{gelfandDiscriminantsResultantsMultidimensional2008}.
Moreover, we can use other kinds of complexes to construct smaller
matrices from which to compute the resultant.
In the best case scenario, the complex will involve only two non-zero
modules and so its determinant will agree with the determinant of the
maps between these modules.
In this case, we will be able to construct \emph{determinantal
  formulas}, that is, maps whose determinant is exactly the resultant,
i.e., the extra factor $E$ is a non-zero constant independent of the
specific system.

Following \cite{gelfandDiscriminantsResultantsMultidimensional2008},
to construct the resultant as the determinant of a complex, we use
sheaf cohomology.
Let $X$ be the normal projective toric variety constructed from the
polytope $P$ detailed before and consider the nef line bundles
$O_X(D_0),\dots,O_X(D_n)$.
Consider global sections $\f_0,\dots,\f_n$ such that $\f_i \in H^0(X,O_X(D_i))$.
If $(\f_0,\dots,\f_n)$ has no common zeros on $X$, the Koszul complex
of sheaves $\mathcal{K}_\bullet$ associated with the Sylvester map,
defined locally at $U \subset X$ as
$\mathtt{Sylv}(g_0,\dots,g_n) \mapsto \sum_i g_i \, \f_i\!\mid_{U}$,
is exact; {\small
 \begin{multline} \mathcal{K_\bullet} : 
   0 \rightarrow O_X(- \sum_i D_i) \rightarrow \bigoplus_i O_X(-
   \sum_{j \neq i} D_j) \rightarrow \dots \\ \dots
   \rightarrow \bigoplus_{j \neq i} O_X(-D_i - D_j)
   \rightarrow \bigoplus_{j} O_X(-D_j) \xrightarrow{\mathtt{Sylv}} O_X
   \rightarrow 0,
 \end{multline}
 }

 \noindent
 where $O_X(D - B) = O_X(D) \otimes O_X(B)^{-1}$, for any two Cartier
 divisors $D$ and $B$.
 We will transform this complex of sheaves into a complex of vector
 spaces by considering its sheaf cohomology. 
 Sheaf cohomology of line bundles on the toric variety $X$ is a
 well-understood subject; for more details, see
 \cite[Chp.~9]{coxToricVarieties2011} and
 \cite{altmann2020displaying}.

 Observe that, for any Cartier divisor $D$, the twisted complex
 $\mathcal{K}_\bullet \otimes O_X(D)$ is exact, because $O_X(D)$ is
 locally invertible. Hence, we consider the exact complex
 $\mathcal{K}_\bullet \otimes O_X(\sum_i D_i)$. Every module in the
 twisted complex is a direct sum of sheaves of the form
 $O_X(\sum_{i \in I} D_i)$, for some $I \subset \{0,\dots,n\}$, so
 they are all nef. Hence, all the higher sheaf cohomologies of the
 modules in $\mathcal{K}_\bullet \otimes O_X(\sum_i D_i)$ vanish
 \cite[Thm.~9.2.3]{coxToricVarieties2011} and so, the associated
 complex of global sections is exact
 \cite[Lem.~2.2.4]{gelfandDiscriminantsResultantsMultidimensional2008},
 {\small
 \begin{multline} H^0(X,\mathcal{K_\bullet} \otimes O_X(\sum_i D_i)) : 
   0 \rightarrow H^0(X,O_X) \rightarrow \dots \\ \dots
   \rightarrow \bigoplus_{j} H^0(X,O_X(\sum_{i \neq j} D_i)) \xrightarrow{\mathtt{Sylv}} H^0(X,O_X(\sum_i D_i))
   \rightarrow 0.
 \end{multline}
 }

 \noindent
 The global sections of the line bundle $O_X(\sum_{i \in I} D_i)$, for
 $I \! \subset \! \{0,\dots,n\}$, correspond to the sparse polynomials with
 Newton polytope equal to (or contained in) $\sum_{i \in I}
 P_i$. Hence, the last map of the exact complex of global sections is 
 the Sylvester map $(g_0,\dots,g_n) \mapsto \sum_i g_i \, \f_i$, where
 $g_i$ is a sparse polynomial such that
 $\NP(g_i) \subset \sum_{j \neq i} P_j$. Canny and Emiris' matrix is a
 square submatrix of the matrix associated with this Sylvester
 map\footnote{However, it is not a submatrix of maximal size; see
   \cite{massriSolvingSparseSystem2016} for more details.}.
 Using this construction, we can also recover a matrix related to
 Macaulay's construction. In this case, we consider $X = \P^n$ and so,
 its Picard group is $\Z$. We identify $D_i$ with the degree of $\f_i$
 and consider the complex
 $\mathcal{K_\bullet} \otimes O_X(\sum_i d_i - n)$. By Serre's
 vanishing theorem, the invertible sheaves appearing in this complex have no
 higher sheaf cohomologies and so, we can consider the exact complex
 of vector spaces
 $H^0(X,\mathcal{K_\bullet} \otimes O_X(\sum_i d_i - n))$. The last
 map of this complex is the Sylvester map
 $(g_0,\dots,g_n) \mapsto \sum_i g_i \, f_i$, where $g_i$ has degree
 $\sum_{j \neq i} \deg(f_j)$. Macaulay's matrix is a maximal square
 submatrix of this map.

 We can also use this construction to understand why the Sylvester matrix
 gives us a determinantal formula. In this case, we have that $n = 1$
 and $X = \P^1$. The aforementioned complex reduces to
 
 \noindent \;\;$ 0 \rightarrow \C[x_0,x_1]_{d_1-1} \oplus \C[x_0,x_1]_{d_0-1}
 \xrightarrow{\mathtt{Sylv}} \C[x_0,x_1]_{d_0+d_1-1} \rightarrow 0,$
 \smallskip
 
 \noindent
 where $d_i = \deg(f_i)$.
 The exactness of the Sylvester map determines the exactness of the
 full complex and agrees with Sylvester's construction. Hence, there
 is no extra-factor in his construction.
 \footnote{Beware that the arguments at our exposition are not
   formal. To prove that the determinant of the complex is the
   resultant we need to consider the Koszul (or, more generally, the
   Weyman) complex not over $X$, but over $X \times \P^{\bm{A}}$,
   i.e., the construction has to be done using systems with generic
   (symbolic) coefficients; see
   \cite{gelfandDiscriminantsResultantsMultidimensional2008}. }

 In the general case, Weyman proposed an approach to derive from
 $\mathcal{K}_\bullet$ an exact complex whose modules are direct sums
 of sheaves cohomologies of the form $H^i(X,O_X(B))$ and such that the
 maps only depend on the coefficients of $(\f_0,\dots,\f_n)$
 \cite{weymanCohomologyVectorBundles2003}.
 By twisting Weyman's complex by certain line bundles and
 studying the vanishing of the sheaf cohomologies appearing in it,
 several authors constructed smaller formulas for computing the sparse
 resultant, e.g.
 \cite{weyman_determinantal_1994,dandrea_macaulay_2005,dickenstein2003multihomogeneous}.
 Moreover, by studying the case when the Weyman complex only involves two
 non-zero modules, this construction was used to derive determinantal
 formulas for the resultant of multihomogeneous systems, e.g.,
 \cite{weyman_determinantal_1994,EmiMan-mhomo-jsc-12,benderBilinearSystemsTwo2018,buse_matrix_2020,emiris_multilinear_2021,benderKoszulTypeDeterminantalFormulas2021}.
 In what follows, we
 present an example taken from
 \cite[Ex.~5.3]{EmiMan-mhomo-jsc-12} of a determinantal formula
 obtained from Weyman's complex which is not a Sylvester formula.
 
 \begin{example}[{Koszul-type formula}] \label{ex:koszForm} We
   consider bilinear forms
   $f_0,f_1,f_2 \in \C[x_0,x_1]_1 \otimes \C[y_0,y_1]_1$. Their
   resultant vanishes if and only if they have a common solution on
   $\P^1 \times \P^1$;
   \begin{align} \label{ex:sysBil}
     \begin{cases}
       f_0 = (a_{0,0} \, x_{0} + a_{1,0} \, x_{1}) \, y_0 +
       (a_{0,1} \, x_{0}  + a_{1,1} \,  x_{1} )  \, y_1 \\
       f_1 = (b_{0,0} \, x_{0} + b_{1,0} \, x_{1}) \, y_0 +
       (b_{0,1} \, x_{0}  + b_{1,1} \,  x_{1} )  \, y_1 \\
       f_2 = (c_{0,0} \, x_{0} + c_{1,0} \, x_{1}) \, y_0 +
       (c_{0,1} \, x_{0}  + c_{1,1} \,  x_{1} )  \, y_1.
     \end{cases}
   \end{align}
   To construct a determinantal formula, we consider the bilinear map
   \begin{align*}
   \star : \C[y_0,y_1]_1 \times (\C[x_0,x_1]_1 \otimes \C[y_0,y_1]_1) \rightarrow \C[x_0,x_1]_1 \\
    y_i \star (x_k \, y_j) := \left\{ \begin{array}{c c} x_j & \text{if } i =
       j, \\ 0 & \text{otherwise}.\end{array}\right. 
   \end{align*}
   The determinant of
   $\phi : (\C[y_0,y_1]_1)^{3} \rightarrow (\C[x_0,x_1]_1)^{3}$ is the
   resultant of the system;
   $$y_i \, \bm{e}_j \mapsto \phi(y_i \, \bm{e}_j) :=
   (y_i \star f_{J_2}) \, \bm{e}_{J_1}
   -
   (y_i \star f_{J_1}) \, \bm{e}_{J_2}, \quad (J = \{0,1,2\} \setminus \{j\})
   $$

   \noindent This formula is
   not a Sylvester formula, but a Koszul one.
   {\small
\begin{align} \label{eq:detForm}
  \begin{array}{c || c c c c | c c}
    & x_{0} \, e_0  & x_{1} \, e_0 & x_{1} \, e_2 & x_{1} \, e_1 & x_{0} \, \, e_2 & x_{0} \, e_1 \\ \hline \hline
    y_{0} \, e_0 & 0       & 0        & b_{1,0} &  -c_{1,0} & b_{0,0}  & -c_{0,0} \\
    y_{1} \, e_0 & 0       & 0        & b_{1,1} &  -c_{1,1} & b_{0,1} &  -c_{0,1} \\
    y_{1} \, e_1 & -c_{0,1} & -c_{1,1} & a_{1,1} & 0       & a_{0,1}   &    0       \\
    y_{1} \, e_2 & -b_{0,1} & -b_{1,1} & 0       & a_{1,1} & 0         &    a_{0,1} \\ \hline
    y_{0} \, e_1 & -c_{0,0} & -c_{1,0} & a_{1,0} & 0       & a_{0,0}   &    0       \\
    y_{0} \, e_2 & -b_{0,0} & -b_{1,0} & 0       & a_{1,0} & 0        &       a_{0,0} 
  \end{array}
\end{align}
}
  
\end{example}

\subsection{Solving sparse systems}
In an analogous way to what we can do with the classical (projective)
resultant, we can use the sparse resultant to compute all the
solutions of a sparse square system.
In this section, we will discuss eigenvalue methods to solve
polynomial systems. Other solving strategies as the hidden-variable
approach or U-resultants
\cite[Chp.~3.5]{coxUsingAlgebraicGeometry2005} can be also extended;
see, e.g., \cite{dandreaPoissonFormulaSparse2015}.

The following method was proposed by Emiris and Rege to recover
multiplication maps using the sparse resultant
\cite{emirisMonomialBasesPolynomial1994}.
Their approach generalizes ideas by Auzinger and Stetter for solving
classical homogeneous systems
\cite{auzingerEliminationAlgorithmComputation1988}.
Consider a sparse polynomial system $f_1,\dots,f_n$ such that their
number of solutions on $\T$ is the mixed volume of
$\NP(f_1),\dots,\NP(f_n)$ and this volume is not zero.
Let $f_0$ be a linear form in the variables $x_1,\dots,x_n$ with
Newton polytope $\Delta_n$.
We consider the matrix constructed by Canny and Emiris' algorithm for
the sparse resultant of $(f_0,\dots,f_n)$
\cite{cannyEfficientAlgorithmSparse1993}. This construction will
generate monomial sets $B_0,\dots,B_n,A$ such that
$\sum \# B_i = \#A$, for any $i$ and $x^\beta \in B_i$, for which
$\Supp(x^\beta \, f_i) \in A$, $\#B_0 = \MV(P_1,\dots,P_n)$ and
$B_0 \subset A$. From these sets, we consider the Macaulay matrix $M$
whose rows are indexed by monomials in $A$ and its columns by pairs
$(i,x^\beta)$ where $x^\beta \in B_i$, and where in the position
corresponding to the row $x^\alpha$ and column $(i,x^\beta)$, the
element is the coefficient of the monomial $x^\alpha$ in
$x^\beta \, f_i$.
We reorder $M$ such that the bottom rows correspond to the pairs
$(0,x^\beta)$ and the left-more columns corresponds to the monomials
in $B_0$, and we split the matrix accordingly:

\begin{center}
\resizebox{\linewidth}{!}{
  \begin{minipage}{1.65\linewidth}
    \huge
  \begin{align} \label{ex:splitMac}
   \begin{tabular}{r}
     $
     \{ x^\beta \, f_i : i > 0, x^\beta \in B_i\}
    \left\{ 
     \lefteqn{\phantom{
     \begin{array}{c}  \\[10pt] M_{1,1}  \\[15pt] \hline \end{array}
     }}
     \right.$\\
     $\{x^\beta \, f_0 : x^\alpha \in B_0 \}
     \left\{\lefteqn{\phantom{
     \begin{array}{c}  \\[0pt]  M_{1,1} \\[5pt] \hline \end{array}
     }}\right.$
   \end{tabular}
  \left[\phantom{\begin{matrix}a_0\\ \ddots\\a_0\\b_0\\ \ddots\\b_0 \end{matrix}}\right.\hspace{-1.5em}
  \overbrace{
  \begin{array}{c |}
    \\[10pt]
    \hspace{18pt} M_{1,1} \hspace{15pt} \\[15pt] \hline\hline
    \\[0pt]
    M_{2,1}         \\[5pt]
  \end{array}
    }^{\text{\huge $A \setminus B_0$}}
    \overbrace{
    \begin{array}{| c}
      \\[10pt]
        \hspace{2pt} M_{1,2} \\[15pt] \hline\hline
      \\[0pt]
      M_{2,2} \\[5pt]
    \end{array}
  }^{\text{\huge $B_0$}}
  \hspace{-1.5em}
    \left.\phantom{\begin{matrix}a_0\\ \ddots\\a_0\\b_0\\ \ddots\\b_0 \end{matrix}}\right]\hspace{-1em}
  \end{align}
\end{minipage}
}
\end{center}
\vspace{.25\baselineskip}

Emiris and Rege showed that if the system $(f_1,\dots,f_n)$ is
  generic-enough then the matrix $M_{1,1}$ is invertible and so the
  Schur complement of $M_{2,2}$, i.e. the matrix
  $M_{2,2} - M_{1,2} \, M_{1,1}^{-1} \, M_{2,1}$, is the
  multiplication matrix of $f_0$ in
  $\C[x_1^{\pm},\dots,x_{n}^{\pm}] / \generators{f_1,\dots,f_n}$. The
  latter holds because this matrix allows us to rewrite elements of
  the form $f_0 \, (\sum_{\beta \in B_0} c_\beta \, x^\beta)$ as
  $\sum_{\beta \in B_0} \bar{c}_\beta \, x^\beta + g$, where
  $g \in \generators{f_1,\dots,f_n}$. If the resultant of
  $(f_0,\dots,f_n)$ does not vanish, then the map is invertible, and
  so $B_0$ is a monomial basis of
  $\C[x_1^{\pm},\dots,x_{n}^{\pm}] / \generators{f_1,\dots,f_n}$ as
  its cardinal $\MV(P_1,\dots,P_n)$ agrees with the dimension of the
  quotient ring \cite{pedersen_mixed_1996}.
  The generic assumptions of the aforementioned construction can be
  relaxed
  \cite{rojasSolvingDegenerateSparse1999,telenNumericalRootFinding2019,benderToricEigenvalueMethods2021}
  and the construction can be extended to solve overdetermined sparse systems
  \cite{massriSolvingSparseSystem2016,benderAnotherEigenvalueAlgorithm2021}.
  Recently, special attention has been given to the numerical aspects of
  this approach
  \cite{telenNumericalRootFinding2019,benderToricEigenvalueMethods2021,benderAnotherEigenvalueAlgorithm2021}.
  In general, we can use determinantal formulas to compute matrices
  which are similar to multiplication matrices, i.e., they share
  the same eigenvalues \cite{benderBilinearSystemsTwo2018}.
  In contrast to the Sylvester formulas, the eigenvectors of the
  determinantal formulas are not well-understood, with the exception
  of the Koszul formulas~\cite{benderBilinearSystemsTwo2018}.
  
  \begin{example}[{Solving using determinantal formulas}]
    Following our Example~\ref{ex:koszForm}, we fix
    $f_0,f_1,f_2 \in \C[x_0,x_1]_1 \otimes \C[y_0,y_1]_1$ as in
    \eqref{ex:sysBil} and partition the determinantal formula as in
    \eqref{eq:detForm},
    \[\left[\begin{array}{c | c}
              M_{1,1} & M_{1,2} \\ \hline
              M_{2,1} & M_{2,2}
      \end{array}
      \right]
    \]
    We will show that, if the coordinates $x_0$ and $y_0$ of every
    solution of $(f_1,f_2)$ are different to zero, then we can solve
    $(f_1,f_2)$ by computing the eigenvalues of the Schur complement
    of $M_{2,2}$.

    By assumption, the system $(x_0 \, y_0, f_1,f_2)$ has no
    solution and so the matrix $M_{1,1}$ is invertible as the
    resultant of $(x_0 \, y_0, f_1,f_2)$ is $\det(M_{1,1}) \neq 0$.
    We introduce a new variable $\lambda$ and consider the resultant
    of the system $(f_0 - \lambda \, x_0, \, y_0, f_1,f_2)$. By the
    Poisson formula \cite{dandreaPoissonFormulaSparse2015}, the
    resultant is a polynomial in $\lambda$ whose roots are of the form
    $\frac{f_0}{x_0 \, y_0}$ evaluated at the solutions of
    $(f_1,f_2)$.
    Hence, the determinantal formula for
    $(f_0 - \lambda \, x_0 \, y_0, f_1,f_2)$ given in
    \eqref{eq:detForm}, can be written as
    \[
      \left[
        \begin{array}{c|c}
          M_{1,1} & M_{1,2} \\
          \hline
          M_{2,1} & M_{2,2} - \lambda \, I
        \end{array}
      \right]
    \]
    where $I$ is the $2 \times 2$ identity matrix. 
    Hence, the resultant of $(f_0 - \lambda \, x_0 \, y_0, f_1,f_2)$
    agrees with the determinant of the Schur complement of
    $M_{2,2} - \lambda \, I$. Equivalently, the eigenvalues of the
    Schur complement of $M_{2,2}$ are the evaluations of
    $\frac{f_0}{x_0 \, y_0}$ at the solutions of $(f_1,f_2)$.
\end{example}

\section{\Groebner bases} \label{sec:gbs}
The objective of this section is to compute \gbs for sparse polynomial
systems.
\gbs algorithms usually can incorporate a priori information of the
system to speed up computations, e.g., syzygy module
\cite{faugereNewEfficientAlgorithm2002,ederSurveySignaturebasedAlgorithms2017}
or Hilbert series \cite{traverso1996hilbert}.
The main issue when we compute with sparse polynomials is that they
behave differently from dense systems, and we do not have much prior
information on them.
We will explain how toric geometry can help us understand the
systems better and so, speed up computations.

\subsection{Embedding of a toric variety} \label{sbsec:embToric}
In this section, we will fix polytopes $P_1,\dots,P_m$ and consider
the projective normal toric variety $X$ associated with
$P := P_1+\dots+P_m$ detailed in the previous section. We assume $P$
is full-dimensional.
Given an integer polytope $Q$, we say it is a \emph{$\N$-Minkowski
  summand} of $P$ if there is $k \in \N$ and an integer polytope $S$
such that $Q + S = k \, P$.
To each $\N$-Minkowski summand $Q$ of $P$ there is a nef Cartier
divisor $D_Q$ such that we can identify the global sections of the nef
line bundle $O_X(D_Q)$ with the polynomials whose Newton polytope is equal
to (or contained in) $Q$, as in \eqref{eq:globalSec}
\cite[Chp.~6]{coxToricVarieties2011}. Moreover, we have that for two
$\N$-Minkowski summands $Q,S$, the global sections of $O_X(D_Q + D_S)$
correspond to the polynomials with Newton polytope $Q + S$.
Given $f_i$ with Newton polytope contained in $P_i$, we will
\emph{homogenize} $f_i$ and write it as
$\f_i \in H^0(X,O_X(D_{P_i}))$.
The common zeros of $\f_1,\dots,\f_m$ will determine a subscheme of
$X$, that we denote by $Y$. The subscheme $Y \cap \T$ agrees with the
subscheme defined by the original $f_1,\dots,f_m$ on $\T$.
Particularly, when $Y \subset \T$, $Y$ is zero-dimensional and we can
study the solutions of $f_1,\dots,f_m$ on $\T$ by studying $Y$.
\begin{example}
  [Intersection theory for toric varieties]
  If $m = n$ and $Y$ is zero-dimensional, the mixed volume
  $\MV(P_1,\dots,P_n)$ is actually the intersection number of
  $D_{P_1},\dots,D_{P_n}$; see
\cite[Chp.~5]{fultonIntroductionToricVarieties1993}.
Hence, the number of solutions of $f_1,\dots,f_n$ is bounded by the
degree of $Y$, and so by this mixed volume. Moreover, generically, the
system $(\f_1,\dots,\f_n)$ has no solutions on $X \setminus \T$ and so
the BKK bound is tight. In general, the BKK bound is tight when we
count the \emph{solutions at infinity}.
\end{example}

In what follows, we assume that there are \emph{$\N$-Minkowski
  summands} $Q_1,\dots,Q_r$ of $P$ and vectors $d_1,\dots,d_m \in \Z^r$
such that $P_i = \sum_k d_{i,k} \, Q_k$.
We consider the $\N^r$-graded algebra given by
$$R^h := \bigoplus_{b \in \N^r}
H^0\left(X,O_X\left(\sum_k b_{k} D_{Q_k}\right)\right)
=
\bigoplus_{b \in \N^r} \!\!\!\!\!\!\!\!
\bigoplus_{\substack{\phantom{A} \\ \alpha \in
    (\sum_k b_{k} \, Q_k) \cap \Z^n}}
\!\!\!\!\!\!\!\!
\C \, x^{(\alpha,b)}.$$

\noindent The Multi-Proj of $R^h$ is the embedding of $X$ in a (weighted)
multi-projective space.

Hence, given polynomials $f_1,\dots,f_m$ with Newton polytopes
$P_1,\dots,P_m$, we can homogenize them as $(\f_1,\dots,\f_m)$ such
that, for each $i$, $\f_i \in S^h_{d_i}$.
If the polynomials $f_1,\dots,f_m$ are generic enough, we can predict
the properties of $(\f_1,\dots,\f_m)$ on $R^h$. For this reason, we
will compute \gbs over $R^h$ and use them to compute \gbs over other
rings.
To formalize this, we introduce the notion of \gbs over
pointed affine semigroup algebras.

\subsection{\gbs for semigroup algebras}
In this section, we will only consider affine semigroups with
identity, that is, a semigroup $(S,+)$ with identity $0$, isomorphic
to a finitely-generated submonoid of $\Z^k$.  We say that $(S,+)$ is
pointed if $\alpha \in S \setminus \{0\}$, $-\alpha \not\in S$, where
$-\alpha$ is defined in the smallest group containing $S$.
Given a semigroup, we define the semigroup algebra $\C[S]$ as the
monomial $\C$-algebra generated by the monomials
$\{x^{\alpha} : \alpha \in S\}$.

If $(S,+)$ is a pointed semigroup, it is possible to define a monomial
ordering $<$ for $\C[S]$; see
\cite[Def.~3.1]{faugereSparseGrObner2014a}. For example, if
$S \subset \N^s$, for some $s \in \N$, then any monomial ordering for
the standard algebra $\C[x_1,\dots,x_s]$ induces a monomial ordering
on $\C[S]$.
Hence, we can define \gbs for ideals in $\C[S]$ using the standard
definition, i.e., $G \subset \C[S]$ is a \gb for an ideal
$I \subset \C[S]$ \wrt a monomial ordering $<$ if $G \subset I$ and,
for every $f \in I$, there is $g \in G$ and $x^\alpha \in \C[S]$ such
that $x^\alpha \, \mathtt{LM}_<(g) = \mathtt{LM}_<(f)$, where
$LM_<(\cdot)$ is the leading monomial in $\C[S]$
\wrt $<$.

Given $\alpha \in \R^n \times \R^r$, let $\pi_1,\pi_2$ be its
projections onto $\R^n$ and $\R^r$, respectively.
Clearly, as $S$ is a semigroup, $\pi_1(S)$ and $\pi_2(S)$ are also
semigroups. In what follows, we assume that, for the affine pointed
semigroup $S$, it holds that $\pi_1(S)$ and $\pi_2(S)$ are also an
affine pointed semigroups and the preimage of each
$\gamma \in \pi_2(S)$ is finite.
Somehow, we will think about $\pi_2(S)$ as a
grading of $S$.
In this case, we say that a monomial ordering $<$ for $S$ is
\emph{graded} if there is another monomial ordering $<_1$ for
$\pi_1(S)$ and $<_2$ for $\pi_2(S)$ such that
$$
x^\alpha < x^\beta \iff
    \left\{
      \begin{array}{l}
        x^{\pi_2(\alpha)} <_2 x^{\pi_2(\beta)} \text{ or }\\
        x^{\pi_2(\alpha)} = x^{\pi_2(\beta)} \text{ and }
        x^{\pi_1(\alpha)} <_1 x^{\pi_1(\beta)}
      \end{array}
    \right.
    $$

   The algebra $R^h$ defined in Subsection~\ref{sbsec:embToric} corresponds
    to the semigroup algebra associated with the semigroup
    $$S^h := \left\{(\alpha, b) \in \Z^n \times \Z^r : \alpha \in \left(\sum_i b_i Q_i\right) \cap \Z^n\right\}.$$

    \noindent
    If $0$ is a vertex of each $Q_i$ and of $\sum_i Q_i$, we can see
    that $S^h$ satisfies the assumptions on the projections $\pi_1$ and
    $\pi_2$ detailed above.
    As the solution set of $(f_1,\dots,f_m)$ over $\T$ does not change
    if we multiply the polynomials by monomials, equivalently, if we
    translate their Newton polytopes, we will assume with no loss of
    generality than $0$ is a vertex of the aforementioned polytopes. 
    Therefore, we consider the semigroup algebra
    $$R := \C[\pi_1(S^h)] \subset \C[x_1^{\pm},\dots,x_{n}^{\pm}].$$

    \noindent We define $\pi$ as the homomorphism between $R^h$ and
    $R$ taking $x^\alpha \mapsto \pi(x^\alpha) :=
    x^{\pi_1(\alpha)}$. This morphism acts as a
    \emph{dehomogenization} and it relates to graded orders as follow.
    Given a Gr\"obner basis $G$ of an ideal of $I \subset R^h$ \wrt a
    graded order $<$, the set $\pi(G)$ is a \gb for $\pi(I) \subset R$
    \wrt $<_1$ \cite[Prop.~3.5]{faugereSparseGrObner2014a}.

\subsection{Computing \gbs}

In this section we discuss an algorithm to compute \gbs for an ideal
$\generators{f_1,\dots,f_r} \subset R$ \wrt a monomial ordering
$<_1$. For this, we will compute \gbs for
$\generators{\f_1,\dots,\f_r} \subset R^h$ \wrt a graded monomial
ordering $<$ associated with $<_1$. In what follows, we ignore the
subindex from the ordering $<_1$.

For each multidegree $b \in \N^r$, we will construct a Macaulay matrix
$M_b$ where the columns are indexed by the monomials of degree $b$ in
$R^h$, the rows by the pairs $(i,x^\beta)$, where $x^\beta$ is a
monomial of degree $b - d_i$. The rows are sorted in decreasing order
\wrt $<$ and the columns are sorted in decreasing order in the following
way: if $i > j$, or $i = j$ and
$x^\alpha > x^\beta$, then $(i,x^\alpha) < (j,x^\beta)$.
The element of the matrix $M_b$ of column index $x^\alpha$, and row
$(i,x^\beta)$ is the coefficient of the monomial $x^\alpha$ in
$x^\beta \f_i$ .
As Lazard realized \cite{lazard_grobner-bases_1983}, if we perform
Gaussian elimination with no pivoting, we obtain new rows representing
elements in the \gb of $\generators{\f_1,\dots,\f_m}$ of degree
$b$. As the \gb is finite, if we perform this computation for
sufficiently many different degrees, we recover the complete \gb.

A classical bottleneck of \gbs computations are reductions to zero. In
the sketched algorithm, these reductions to zero correspond to the
rows reducing to zero after performing Gaussian elimination. An
established way of speeding up computations is to avoid these
reductions using the F5 criterion
\cite{faugereNewEfficientAlgorithm2002,ederSurveySignaturebasedAlgorithms2017}.
In this context, the F5 criterion is called Matrix-F5
\cite[Sec.~3]{ederSurveySignaturebasedAlgorithms2017} and translates into
skipping some rows from the construction of the matrix.
The F5 criterion tells us that the row indexed by $(i,x^\beta)$ will
reduce to zero after performing Gaussian elimination if and only if
$x^\beta$ is the leading monomial of a polynomial of degree
${b - d_i}$ in the colon ideal
$(\generators{\f_1,\dots,\f_{i-1}} : f_i)$, i.e.,
$x^\beta \in LM((\generators{\f_1,\dots,\f_{i-1}} : f_i))_{b - d_i}$.
We can organize our computation in such a way that we can recover from
it $LM(\generators{\f_1,\dots,\f_{i-1}})_{b - d_i}$
\cite[Sec.~3]{ederSurveySignaturebasedAlgorithms2017}. Hence, when
these two sets agree, we can skip every reduction to zero, that is,
the F5 criterion is \emph{optimal}.
A classical setting where this happens is when $f_i$ is not a zero
divisor in $R^h / \generators{\f_1,\dots,\f_{i-1}}$. Hence, when
$(\f_1,\dots,\f_m)$ is a regular sequence, i.e., when the previous condition
holds for every $i$, then we can avoid every reduction to zero.
In the unmixed case, we have that $r = 1$ and the ring $R^h$ is
$\N$-graded. It can be shown that this is a Cohen-Macaulay ring and
that, if $m \leq n$, a generic unmixed sparse system
$(\f_1,\dots,\f_m)$ forms a regular sequence on $R^h$
\cite{hochsterRingsInvariantsTori1972}.
This last fact was exploited by Faug\`ere, Spaenlehauer, and Svartz
\cite{faugereSparseGrObner2014a} to compute \gbs for unmixed systems.
However, in the mixed case, this is not true anymore, that is, our
system $(\f_1,\dots,\f_m)$ usually does not form a regular sequence on
$R^h$.
To solve this problem, together with Faug\`ere and Tsigaridas
\cite{bender_towards_2018,bender_groebner_2019}, we observed that the
F5 criterion is optimal at \emph{certain degree} $b$ whenever
$(\generators{\f_1,\dots,\f_{i-1}} : f_i)_{b - d_i} =
\generators{\f_1,\dots,\f_{i-1}}_{b - d_i}$, for each $i$.
This condition can be enforced by the vanishing of the first homology
of the Koszul complex of $(\f_1,\dots,\f_i)$ at degree $b$, for each
$i$.
We can prove that, if the subscheme defined by $(\f_1,\dots,\f_i)$ on
$X$ is a complete intersection, then the first homology of the Koszul
complex of $(\f_1,\dots,\f_i)$ vanishes at any degree
$b \geq \sum_{j \leq i} d_i$. Moreover, this bound can be improved for
special systems. For example, for multihomogeneous systems over
$\Pr^{n_1} \times \dots \times \Pr^{n_r}$, we have that
$b \geq \sum_{j \leq i} d_i - (n_1,\dots,n_r)$
suffices
\cite{bender_towards_2018}
\footnote{A spectre is haunting
  this argument - a spectral sequence associated to the Koszul complex
  of sheaves mentioned in the previous section. Normal projective
  toric varieties are Cohen-Macaulay, so if $(\f_1,\dots,\f_i)$
  defines a complete intersection $Y$ on $X$, the Koszul complex gives
  a resolution for the structure sheaf $O_Y$.
  In this case, we can transform this complex into a resolution of
  $R^h / \generators{f_1,\dots,f_{i-1}}$ at degree $b \in \N^r$, if
  for each $I \subseteq \{1,\dots,i\}$,
  $H^t(X,O_X(\sum_j (b_j - \sum_{k \in I} d_{j,k}) D_{Q_j})) = 0$, for
  $t > 0$.
  It is easy to construct a priori bounds for $b$ as $\sum_j d_i$
  using vanishing theorems for the sheaf cohomology of line bundles
  over toric varieties; see, e.g.,
  \cite[Sec.~4]{benderToricEigenvalueMethods2021}.
  In our papers \cite{bender_towards_2018,bender_groebner_2019}, we
  skipped this discussion by defining the concept of \emph{Koszul
    regular systems} as the systems such that, for every subsystem,
  the first homology of its associated Koszul complex vanishes at
  certain degrees, which we can predict from the previous analysis.
}.
Hence, we can predict all the reductions to zero appearing at
``big-enough'' degree. Moreover, if our objective is to only compute
\gbs over $R$, we will just avoid the degrees at which we cannot predict
every reduction.

\begin{remark}
  This strategy generalizes partially previous work on computing \gbs
  for special sparse systems such as bilinear 
  \cite{faugere_grobner_2011} and weighted homogeneous systems
  \cite{faugere_complexity_2016}.
\end{remark}

To derive complexity bounds, we need to construct a bound of the
maximal degree of an element in a \gbs of
$\generators{\f_1,\dots,\f_m}$. However, the monomial orderings
defined over semigroup algebras, in general, does not have the
dehomogenization property of GRevLex
\cite[Ex.~7.2.1]{bender_thesis_2019}. Moreover, we cannot talk about
generic coordinates for the solutions, as random change of coordinates
destroys the sparsity. Hence, we cannot use strategies as
\cite{bayer_criterion_1987} to construct bounds for the maximal
degrees of the elements in the \gbs. As we will show in the next
section, we can work around this problem in the zero-dimensional case,
i.e., when the subscheme associated to
$\generators{\f_1,\dots,\f_m}$ has dimension zero.

\subsection{Zero-dimensional systems}
In this section, we will construct bounds for computing \gbs of
zero-dimensional ideals. 
Classically, to compute \gbs for zero-dimensional systems, we first
calculate a \gb \wrt GRevLex and then we use FGLM to recover the \gb
\wrt the ordering we want. We do so because, \wrt GRevLex, the \gb
involves elements of smaller degrees, for which we may find upper
bounds. Generally, the maximal degree of an element in a \gb of a
zero-dimensional ideal is given by the \emph{Castelnuovo-Mumford
  regularity} of its homogenization
\cite[Cor. 3]{chardin_bounds_2003}.
In the semigroup setting, where there usually does not exist a GRevLex
ordering, the maximal degree of an element in the \gb could be higher
than the regularity; see \cite[Sec.~8.3.1]{bender_thesis_2019}.
To work around this problem, in
\cite{bender_towards_2018,bender_groebner_2019}, we presented a way to
recover the multiplication maps by truncating our computation of \gbs
at a certain degree. By doing so, we bounded the complexity of our
computation.
In what follows, we explain the idea behind this strategy in the
general case of mixed sparse polynomial systems and refer the reader
to \cite[Chp.~8]{bender_thesis_2019} for improvements in special
cases.

Following the notation from the previous section, we consider an
affine sparse system $(f_1,\dots,f_m)$ such that the subscheme $Y$
associated with $\generators{\f_1,\dots,\f_m} \subset R^h$ has only
solutions on $\T$ (in particular, $Y$ is finite as $X$ is
compact). That is, it has no solutions at infinity.
We assume without loss of generality that
$1,x_1,\dots,x_n \in R$.\footnote{If the monomials do not belong to
  $R$, we change the toric variety $X$ by another one associated with
  $P + \Delta_n$. As $Y \subset \T$, this change does not affect $Y$
  and adds a degree $d_0$ to $R^h$ as wanted; see
  \cite[Sec.~4]{bender_groebner_2019}.}
In this case, we can prove \cite[Lem.~4.11]{bender_groebner_2019},
$$R / \pi(\generators{\f_1,\dots,\f_m}) \simeq \C[x_1,\dots,x_n] /
(\generators{f_1,\dots,f_n} : \generators{\Pi_i x_i}^\infty).$$

\noindent
Our strategy will be to perform FGLM using multiplication maps over
$R / \pi(\generators{\f_1,\dots,\f_m})$ to recover the \gb of
$\generators{f_1,\dots,f_m} : \generators{\Pi_i x_i}^\infty$ in
$\C[x_1,\dots,x_n]$. In what follows, we focus on computing
these multiplication maps by adapting Emiris and Rege \cite{emirisMonomialBasesPolynomial1994}.

Consider $b = \sum_i d_i$, where $d_i$ is the multidegree of $\f_i$. Let
$d_0$ be a multidegree such that
$\Delta_n \cap \N^n \subset \pi(R^h_{d_0})$, that is, after
dehomogenization, we can obtain a linear form in $R$.
At the degrees $b$ and $b + d_0$, the ideal
$R^h / \generators{\f_1,\dots,\f_m}$ is a vector space of dimension
$\delta$ equal to the number of points in $Y$, counting multiplicities
\cite[Thm.~3]{massriSolvingSparseSystem2016}. Let $B_0$ be the set of
monomials of degree $d$ which are not leading monomials in
$LM(\generators{\f_1,\dots,\f_{m}})$. The cardinality of $B_0$ is
$\delta$. Moreover, also by
\cite[Thm.~3]{massriSolvingSparseSystem2016}, as there are no
solutions outside $\T$, for any monomial $x^\alpha \in R^h_{d_0}$,
$R^h_{b+d_0} = \generators{\f_1,\dots,\f_m,x^\alpha}_{b + d_0}$, and
using the F5 criterion, it is easy to see that the monomial set
$\{x^\alpha \, x^\beta : x^\beta \in B_0\}$ forms a basis of the
vector space $(R^h / \generators{\f_1,\dots,\f_m})_{b + d_0}$.
Let $x^{\alpha_0}$ be the monomial of degree $b+d_0$ such that
$\pi(x^{\alpha_0}) = 1$.
Then, we can use our Macaulay matrix at degree $b + d_0$ related to
$(\f_1,\dots,\f_m)$ to rewrite any product $x^\beta \, f_0$, where
$f_0 \in R^h_{d_0}$ and $x^\beta \in B_0$, as a linear combination of
monomials in $\{x^\alpha \, x^\beta : x^\beta \in B_0\}$.
By dehomogenizing this relation, we recover the multiplication map of
$\pi(f_0)$ over $R / \generators{\pi(\f_1),\dots,\pi(\f_m)}$,
\wrt the basis $\pi(B_0)$. As the monomials $1,x_1,\dots,x_n$ belong
to $\pi(B_0)$, we can recover every multiplication map in
$\C[x_1,\dots,x_n] / (\generators{f_1,\dots,f_n} : \generators{\Pi_i
  x_i}^\infty)$.

\begin{acks}
  This manuscript is dedicated to the memory of Agnes Szanto. Thank
  you very much, Agnes, for your selfless help and support.

  I would like to thank Christopher Borger, Jean-Charles Faug\`ere,
  Angelos Mantzaflaris, Pierre-Jean Spaenlehauer, Elias Tsigaridas,
  and Simon Telen, for our joint work where I learned a lot about what
  is detailed in this manuscript.
  Also, to Laurent Bus\'e and Marc Chardin, for teaching me how to use
  local cohomology and spectral sequences to study the
  Castelnuovo-Mumford regularity.
  I am grateful to Fatima Abu Salem and Amir Hashemi for their
  several comments and corrections on this manuscript.
  
  This work was funded by the ERC under the European's
  Horizon 2020 research and innovation programme (grant agreement
  787840).
\end{acks}

\bibliographystyle{ACM-Reference-Format}
\bibliography{biblio}

\end{document}